\documentclass[aps,pra,reprint,amsmath,amssymb,superscriptaddress,longbibliography]{revtex4-2}

\usepackage{color,soul}
\usepackage{graphicx}
\usepackage[colorlinks=true,urlcolor=blue,citecolor=blue,linkcolor=blue,bookmarks=false,pdfstartview={FitH}]{hyperref}

\begin{document}

\title{Anisotropy of Kondo-lattice coherence in momentum space for CeCoIn$_5$}

\author{Mai Ye}
\email{mai.ye@kit.edu}
\altaffiliation[Present address: ]{Institute for Quantum Materials and Technologies, Karlsruhe Institute of Technology, 76021 Karlsruhe, Germany}
\affiliation{Department of Physics and Astronomy, Rutgers University, Piscataway, NJ 08854, USA}
\author{Hsiang-Hsi Kung}
\email{hhseankung@gmail.com}
\altaffiliation[Present address: ]{Stewart Blusson Quantum Matter Institute, University of British Columbia, V6T1Z4 Vancouver, Canada}
\affiliation{Department of Physics and Astronomy, Rutgers University, Piscataway, NJ 08854, USA}
\author{Priscila F. S. Rosa}
\affiliation{Los Alamos National Laboratory, Los Alamos, NM 87545, USA}
\author{Eric D. Bauer}
\affiliation{Los Alamos National Laboratory, Los Alamos, NM 87545, USA}
\author{Kristjan Haule}
\affiliation{Department of Physics and Astronomy, Rutgers University, Piscataway, NJ 08854, USA}
\author{Girsh Blumberg}
\email{girsh@physics.rutgers.edu}
\affiliation{Department of Physics and Astronomy, Rutgers University, Piscataway, NJ 08854, USA}
\affiliation{National Institute of Chemical Physics and Biophysics, 12618 Tallinn, Estonia}

\date{\today}

\begin{abstract}
We study the electronic and phononic excitations of heavy-fermion metal CeCoIn$_5$ by polarization-resolved Raman spectroscopy to explore the Kondo-lattice coherence. 
Below the coherence temperature T*\,=\,45\,K, the continuum of electronic excitations in the XY scattering geometry is suppressed at frequencies below 50\,cm$^{-1}$, whereas the low-frequency continuum in the X'Y' geometry exhibits no change across T*.  
We relate the suppression to the reduced electron-electron scattering rate resulting from the coherence effect. 
The presence of suppression in the XY geometry and absence of it in the X'Y' geometry implies that the $\alpha$ and $\beta$ bands become coherent below T*, whereas the $\gamma$ band remains largely incoherent down to 10\,K. Moreover, two optical phonon modes exhibit anomalies in their temperature dependence of the frequency and linewidth below T*, which results from developing coherent spectral weight near the Fermi level and reduced electron-phonon scattering rate. Our results further support the key role of anisotropic hybridization in CeCoIn$_5$.
\end{abstract}

\maketitle

\section{Introduction\label{sec:Intro}}

The key physics of Kondo-lattice materials is the interaction between itinerant conduction electrons and a lattice of localized electrons~\cite{Hewson1993,Coleman2015Note}. The itinerant conduction electrons may either screen local moments, forming Kondo singlets, or facilitate long-range interactions between local moments, leading to a magnetically-ordered ground state~\cite{Coleman2015Book}. The local moments may also induce an effective interaction between itinerant electrons via the exchange of virtual collective fluctuations. In particular, spin fluctuations have been proposed as a pairing mechanism for heavy-fermion superconductors~\cite{Scalapino2012}.

Among the broad range of open questions in the field, one interesting issue is the evolution from incoherent scattering at high temperatures to a coherent state at low temperatures. In this process, the localized electrons acquire effective intersite coupling and gradually form a heavy-fermion band~\cite{Hewson1993}. Because of the development of coherence between sites, the electron scattering is suppressed and hence the electric resistivity decreases on cooling. Moreover, the collective deconfinement of localized electrons leads to a deviation from Curie behavior for magnetic susceptibility, accompanied by an anomaly in Knight shift measured by nuclear magnetic resonance (NMR)~\cite{Curro2004}. The characteristic temperature marking the change of electric resistivity and magnetic susceptibility is commonly denoted as the coherence temperature T*~\cite{Yang2008}.

The conventional idea that the formation of the heavy-fermion band happens below T*, however, is being challenged by new experimental results. The case of heavy-fermion metal CeCoIn$_5$, a prototypical Kondo-lattice system, serves as an insightful example. This material shows no magnetic ordering at low temperature, but it becomes superconducting below T$_c$=2.3\,K~\cite{Resistivity2001}. The coherence temperature of CeCoIn$_5$ has been identified at T*=45\,K by resistivity~\cite{Resistivity2001}, NMR~\cite{NMR2003}, and inelastic neutron scattering~\cite{Severing2010} measurements. Because of its proximity to a magnetic instability, this compound exhibits strange-metal behavior evidenced, for example, by its linear-in-temperature dependence of resistivity~\cite{Resistivity2001} below about 20\,K. Recent angle-resolved photoemission (ARPES) studies observe finite $f$-electron spectral weight near the Fermi level up to around 200\,K~\cite{ARPES2017,ARPES2020}, well above T*. These ARPES data further suggest that crystal-field (CF) effect influences both the Kondo temperature and the $f$-electron spectral weight near the Fermi level. The development of a spectral gap at temperatures far above the coherence temperature has also been observed in other heavy-fermion compounds, for example Ce$_2$RhIn$_8$~\cite{adriano2015,rodolakis2018}.

To gain more insight into this issue, it is important to examine the temperature dependence of other relevant physical quantities of CeCoIn$_5$ and identify the temperatures at which anomalies occur. Moreover, as the ARPES studies mentioned above as well as earlier experimental~\cite{Pagliuso2002,Severing2015,Rosa2019} and theoretical~\cite{Shim2007,Haule2010} work emphasize the role played by CF effects, it is desired to directly probe the CF excitations and study their evolution with temperature. The electronic configuration of Ce$^{3+}$ ion is 4$f^1$. According to Hund's rules, the sixfold degenerate $^2F_{5/2}$ multiplet has the lowest energy. The tetragonal CF potential at Ce$^{3+}$ sites splits the $^2F_{5/2}$ multiplet into two $\Gamma_7$ and one $\Gamma_6$ doublets.

ARPES, a single-electron spectroscopy, probes fermionic excitations, whereas electronic Raman scattering spectroscopy serves as a complementary method capable of measuring bosonic collective modes and particle-hole excitations. 
Particularly, the electronic Raman spectroscopy is well suited to study CF excitations~\cite{Cardona2000}. 
In this work, we study the electronic and phononic excitations of CeCoIn$_5$ using polarization-resolved Raman spectroscopy. We specifically investigate (i) low-frequency excitations related to conduction electrons, (ii) CF transition related to the $f$ electrons, and (iii) optical phonon modes relevant to the lattice. We find two phenomena to happen only below T*: suppression of the low-frequency electronic continuum and anomalous temperature dependence of the phonon spectral parameters. In particular, the suppression occurs in the XY scattering geometry but not in the X'Y' geometry. By relating the suppression to the reduced electron-electron scattering rate, we show that the coherence process is anisotropic in momentum space, with the $\alpha$ and $\beta$ bands become coherent below T*, whereas the $\gamma$ band remains largely incoherent down to 10\,K.

\section{Experimental\label{sec:Exp}}

Single crystals of CeCoIn$_5$ were grown using the in-flux technique~\cite{Canfield1992,Rosa2018}.
Two samples were used in the Raman study: one had a clean as-grown $xy$ crystallographic plane; the other was cleaved in ambient conditions to expose its $xz$ crystallographic plane.
The sample surfaces were examined under a Nomarski microscope to find a strain-free area.
Raman-scattering measurements were performed in a quasi-back-scattering geometry from the samples mounted in a continuous helium gas flow cryostat.

We used a custom triple-grating spectrometer with 1800\,mm$^{-1}$ master holographic gratings for data acquisition. All data were corrected for the spectral response of the spectrometer.

\begin{table}[t]
\caption{The relationship between the scattering geometries and the symmetry channels. For scattering geometry E$_{i}$E$_{s}$, E$_{i}$ and E$_{s}$ are the polarizations of incident and scattered light; X, Y, X', Y' and Z are the [100], [010], [110], [1$\overline{1}$0] and [001] crystallographic directions. A$_{1g}$, A$_{2g}$, B$_{1g}$, B$_{2g}$ and E$_{g}$ are the irreducible representations of the D$_{4h}$ group.}
\begin{ruledtabular}
\begin{tabular}{ll}
Scattering Geometry&Symmetry Channel (D$_{4h}$)\\
\hline
XX&A$_{1g}$+B$_{1g}$\\
XY&A$_{2g}$+B$_{2g}$\\
X'X'&A$_{1g}$+B$_{2g}$\\
X'Y'&A$_{2g}$+B$_{1g}$\\
XZ&E$_{g}$\\
\end{tabular}
\end{ruledtabular}
\label{tab:Exp1}
\end{table}
Five polarization configurations were employed to probe excitations in different symmetry channels. The relationship between the scattering geometries and the symmetry channels~\cite{Hayes1978} is given in Table\,\ref{tab:Exp1}.

The 752\,nm line from a Kr$^+$ ion laser was used for excitation. Incident light was focused to a 50$\times$50\,$\mu$m$^{2}$ spot. Laser power less than 20\,mW was used. The reported temperature values were corrected for laser heating, with the heating rate estimated to be 0.5\,K$\slash$\,mW\,\footnote{We base the estimate of laser heating at low temperatures on a comparison with other heavy-fermion metals of comparable electrical conductivity~\cite{Resistivity2001}: CeB$_6$~\cite{Ye2019Ce} and YbRu$_2$Ge$_2$~\cite{Ye2019Yb}, for which laser heating with the same experimental setup was established in the range of 0.5-0.75\,K$\slash$\,mW}.

The measured secondary-emission intensity $I(\omega,T)$ is related to the Raman response $\chi''(\omega,T)$ by $I(\omega,T)=[1+n(\omega,T)]\chi''(\omega,T)+L$, where $n$ is the Bose factor, $\omega$ is energy, $T$ is temperature, and $L$ represents photoluminescence which exhibits no frequency and temperature dependence in the narrow spectral window of interest.

\section{Overview\label{sec:OV}}

\begin{figure}
\includegraphics[width=\linewidth]{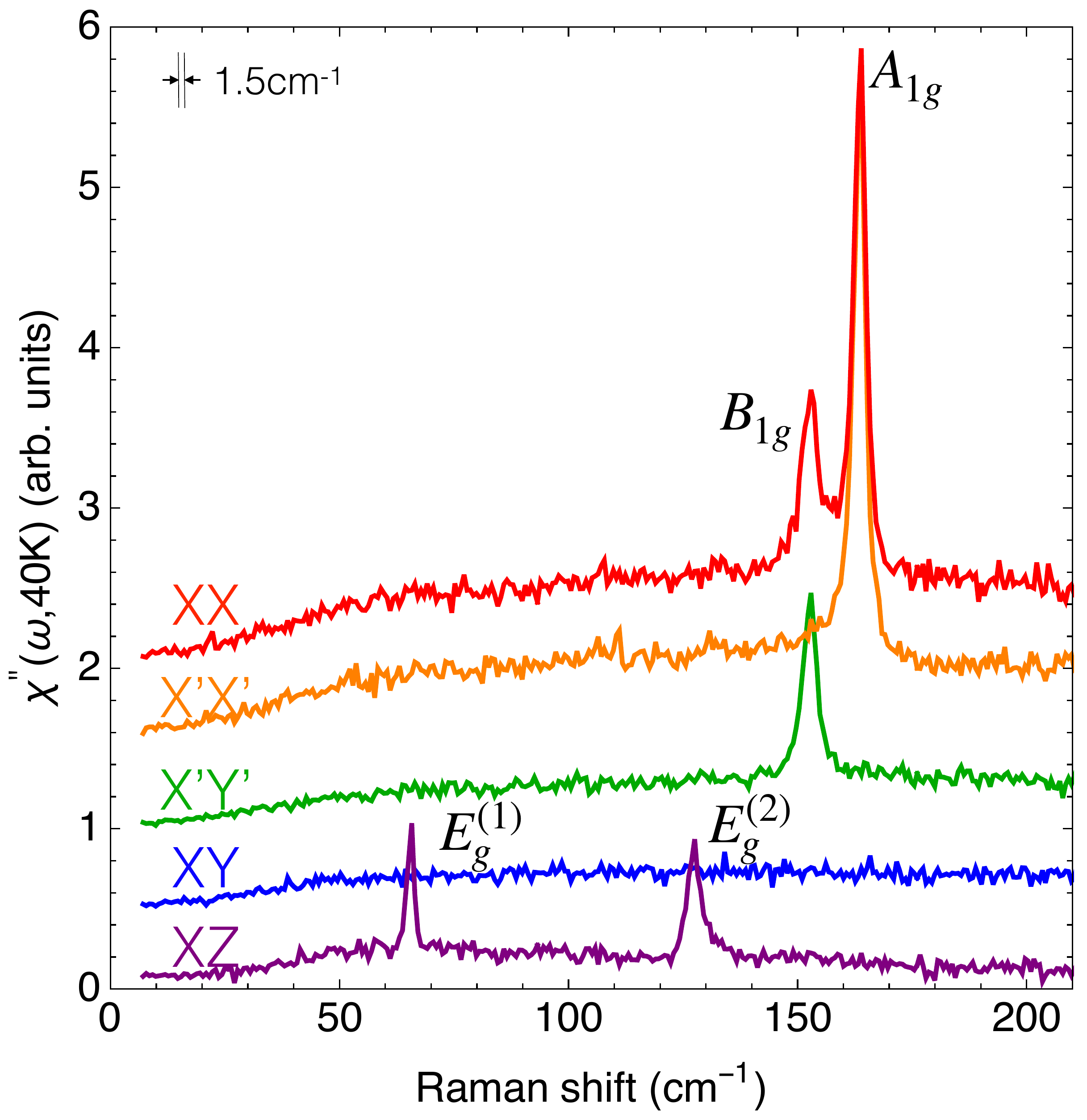}
\label{fig:P}
\caption{Polarization dependence of the Raman spectra measured at 40\,K. The four Raman-active optical phonon modes are labelled by their respective symmetry.}
\end{figure}

In Fig.\,\ref{fig:P} we present the Raman spectra measured at 40\,K in five scattering geometries. 
The broad continuum corresponds to electronic excitations, and the sharp modes are phonons. From group-theoretical analysis, CeCoIn$_5$ has four Raman-active optical phonon modes: $1A_{1g}\oplus 1B_{1g}\oplus 2E_{g}$. The frequency of the $A_{1g}$ phonon mode is consistent with the value reported by a previous Raman study~\cite{Raman2007}.

\section{Electronic Excitations\label{sec:E}}

\begin{figure}
\includegraphics[width=0.95\linewidth]{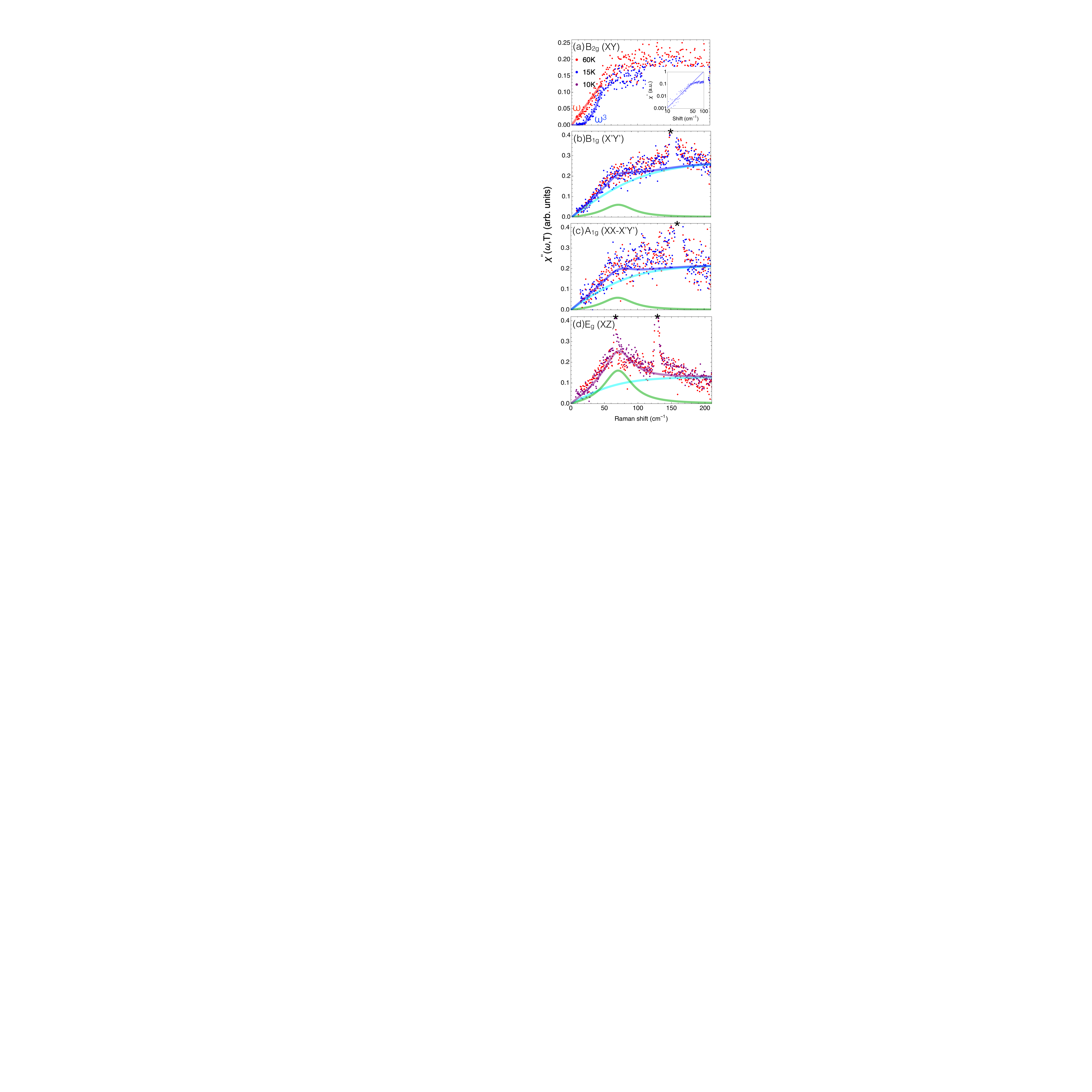}
\caption{\label{fig:E}Low-frequency electronic excitations of CeCoIn$_5$ measured in (a) $B_{2g}$, (b) $B_{1g}$, (c) $A_{1g}$ and (d) $E_{g}$ symmetry channels, obtained by decomposing the Raman spectra measured in different scattering geometries with the help of Table\,\ref{tab:Exp1}. In panel (a), the red line and blue curve represent linear and cubic functions, respectively; the inset is a log-log plot of the 15\,K data compared with a cubic function. In panels (b), (c), and (d), the cyan and green curves show hyperbolic-tangent and Lorentzian functions, respectively; the sum of the cyan and green curves is represented by a blue curve in panels (b) and (c), and by a purple curve in panel (d). The phonon modes, which have been shown in Fig.\,\ref{fig:P}, are labelled by asterisks.}
\end{figure}

In Fig.~\ref{fig:E} we present the low-frequency electronic excitations of CeCoIn$_5$ in four symmetry channels. We measure the spectra in XX, XY, X'Y', and XZ scattering geometries, and we decompose the signal into the $A_{1g}$, $B_{1g}$, $B_{2g}$, and $E_{g}$ symmetry channels following Table\,\ref{tab:Exp1}; in decomposition we assume that the signal in the anti-symmetric $A_{2g}$ channel is weak\,\footnote{Although the $\Gamma_7\rightarrow\Gamma_6$ CF transition at 200\,cm$^{-1}$~\cite{Neutron2004,Severing2010} has $A_{2g}$ component, this CF mode is not resolved in the Raman spectra.}. 
Both a continuum of excitations due to conduction electrons and a mode related to transition between $f$-electron CF states are observed.

The lowest-lying $^2F_{5/2}$ multiplet contains two $\Gamma_7$ and one $\Gamma_6$ doublets. From group-theoretical analyses, the $\Gamma_7\rightarrow\Gamma_7$ transition has $A_{1g}\oplus B_{1g}\oplus E_{g}$ symmetry; the $\Gamma_7\rightarrow\Gamma_6$ transition has $A_{2g}\oplus B_{2g}\oplus E_{g}$ symmetry. 
Because the electronic mode observed at 70\,cm$^{-1}$ contains $A_{1g}$ and $B_{1g}$ components, it corresponds to the $\Gamma_7\rightarrow\Gamma_7$ CF transition. 
The CF ground state, in turn, must be of $\Gamma_7$ symmetry (if $\Gamma_6$ would be the ground state, $\Gamma_7\rightarrow\Gamma_7$ CF transition could not be observed.). 
The assignment of a $\Gamma_7$ ground state and a $\Gamma_7$ excited state at 70\,cm$^{-1}$ is consistent with previous neutron and X-ray measurements~\cite{Neutron2004,Severing2010,Severing2019}. The $\Gamma_7\rightarrow\Gamma_6$ transition has a frequency of 200\,cm$^{-1}$ determined by inelastic neutron scattering~\cite{Neutron2004,Severing2010}, but this mode is not resolved in the Raman spectra. Because for the $\Gamma_7\rightarrow\Gamma_7$ transition the full width at half maximum is comparable to the frequency even at 10\,K [Fig.~\ref{fig:E}(d)], one possible reason for not resolving the $\Gamma_7\rightarrow\Gamma_6$ transition is that the lineshape of this mode is too broad to be distinguished from the electronic continuum.

In the $A_{1g}$, $B_{1g}$ and $E_{g}$ symmetry channels, we model the low-frequency electronic continuum by a phenomenological hyperbolic-tangent function, $\chi^{\prime\prime}(\omega) \sim \tanh (\omega/\omega_0)$, which is linear in frequency for $\omega\ll\omega_0$, and becomes frequency-independent for $\omega\gg\omega_0$. The $\Gamma_7\rightarrow\Gamma_7$ CF transition in these symmetry channels is modeled by a Lorentzian function.

In the $B_{2g}$ symmetry channel, a suppression of the electronic continuum up to around 50\,cm$^{-1}$ happens below the coherence temperature T* [Fig.~\ref{fig:E}(a)]. The Raman response below 50\,cm$^{-1}$ evolves from linear frequency dependence above T* to a cubic-power-law dependence below T*. In contrast, the Raman response remains essentially unchanged at low frequency in the $B_{1g}$ symmetry channel [Fig.~\ref{fig:E}(b)].

The electronic continuum results from strong scattering and large electronic self-energy~\cite{Shastry2018}. Because density fluctuations give rise to a vanishing contribution in the $q\rightarrow 0$ limit as a result of momentum conservation, such continuum is not observed for weakly correlated metals~\cite{Hayes1978}. In strongly correlated systems, the incoherent background of the electronic spectral functions leads to the continuum of electronic excitations.

We therefore relate the suppression in the $B_{2g}$ channel to the coherence effect. The change of the electronic structure, and in turn the development of coherent spectral weight near the Fermi level below T*, reduces the scattering near the Fermi level and hence suppresses the continuum at low frequency. The 50\,cm$^{-1}$ value therefore serves as a characteristic frequency scale for the development of coherence.

\begin{figure}
\includegraphics[width=0.95\linewidth]{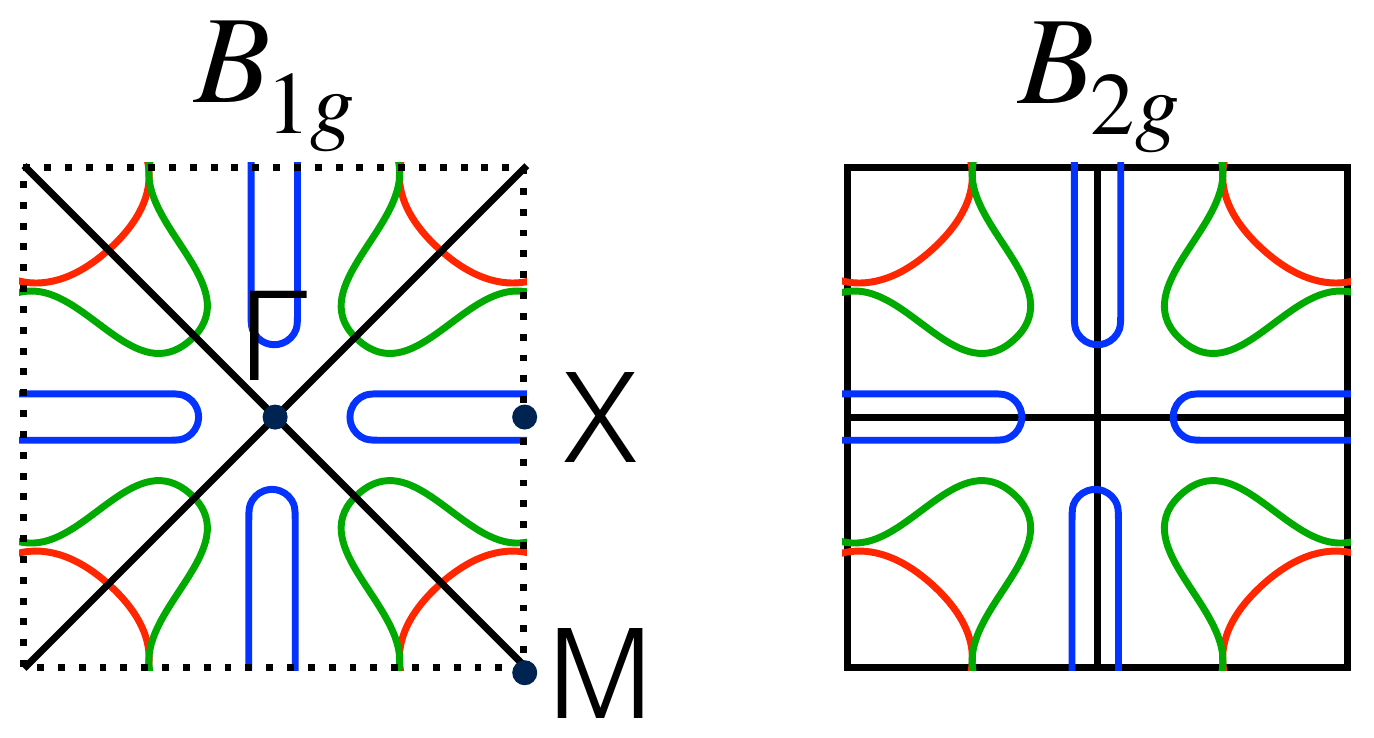}
\caption{\label{fig:BZ}Schematic plots of the nodal lines for the light-scattering vertex in the $B_{1g}$ and $B_{2g}$ symmetry channels. The Brillouin zone is indicated by the dotted black lines, and the solid black lines corresponds to lines along which the light-scattering vertex is zero for the $B_{1g}$ and $B_{2g}$ cases, respectively. The $\alpha$, $\beta$, and $\gamma$ bands~\cite{ARPES2017,ARPES2020,Analytis2021} are shown in red, green, and blue colors, respectively.}
\end{figure}

The different temperature dependence of the low-frequency Raman response in the $B_{1g}$ and $B_{2g}$ symmetry channels further points to momentum dependence of the Kondo-lattice coherence. In these two channels, different regions of the Brillouin zone are selectively probed~\cite{Shastry1990,Devereaux1994,Shastry2018}. Although the light-scattering vertex can be complicated for realistic bands, symmetry properties require that the vertex is zero along the diagonals of the Brillouin zone for the $B_{1g}$ channel, whereas the zero-vertex condition occurs along the boundaries of the Brillouin-zone quadrants for the $B_{2g}$ symmetry channel~\cite{Shastry1990,Devereaux1994} [Fig.~\ref{fig:BZ}]. The suppression of the electronic continuum in the $B_{2g}$ channel but not in the $B_{1g}$ channel therefore implies that the $\alpha$ and $\beta$ bands become coherent below T*, whereas the $\gamma$ band remains largely incoherent.

\section{Phononic Excitations\label{sec:P}}

\begin{figure}[b]
\includegraphics[width=0.95\linewidth]{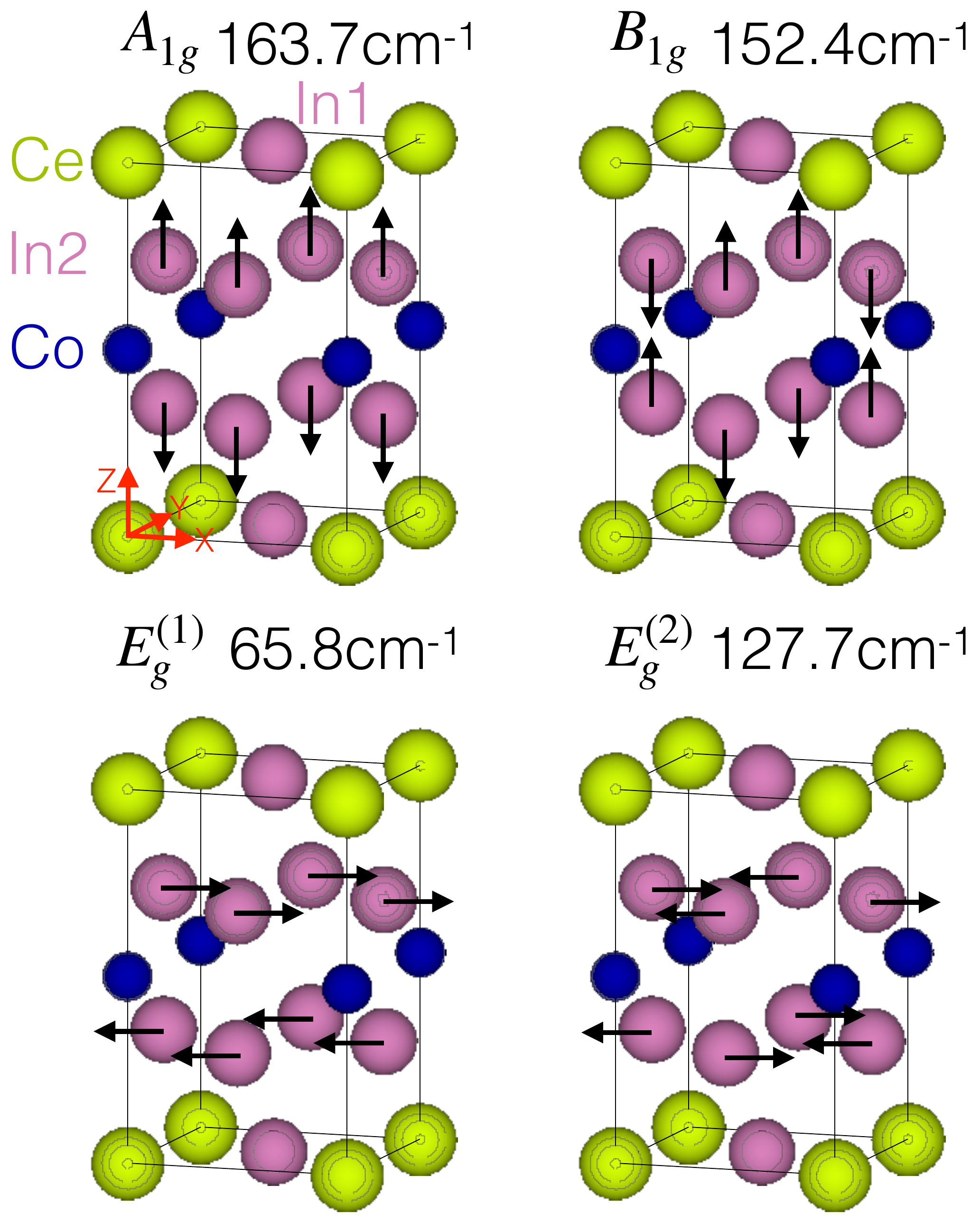}
\caption{\label{fig:Pattern}Schematic vibration patterns of the Raman-active phonon modes. These modes are classified by the irreducible representations of the D$_{4h}$ point group. The frequencies of these modes measured at 5\,K are labelled.}
\end{figure}

The vibration patterns of the phonon modes are provided in Fig.~\ref{fig:Pattern}. As the Ce, Co, and In1 sites are located at inversion centers, only In2 atoms participate in Raman-active phonon vibrations~\footnote{In1 atoms are located in the same layers as Ce atoms; In2 atoms are located in the layers between Ce an Co leyers}. The $A_{1g}$ and $B_{1g}$ modes involve vibrations of In2 atoms along the $z$ direction, while the two $E_{g}$ modes are related to in-plane motion of In2 atoms.

\begin{figure}
\includegraphics[width=0.95\linewidth]{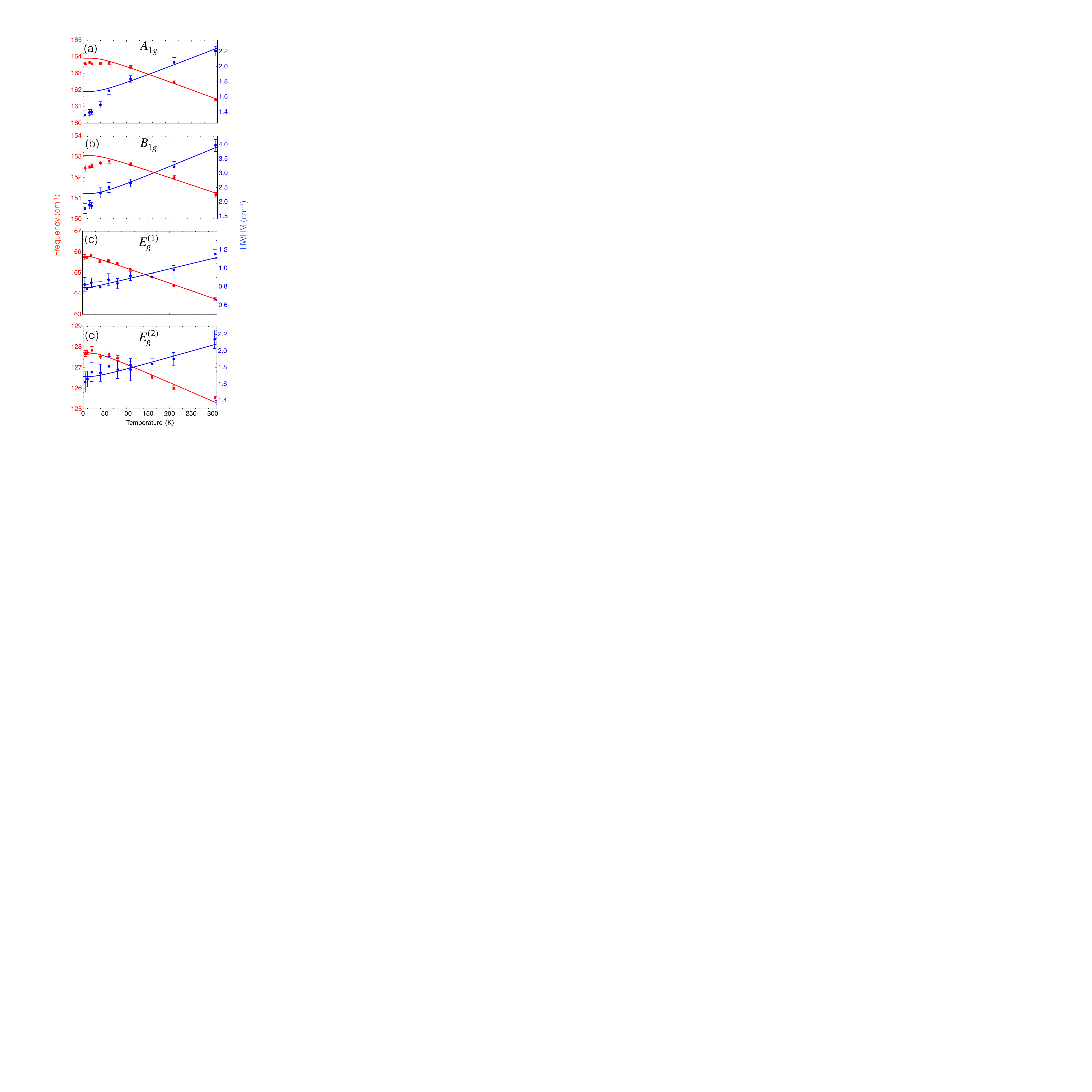}
\caption{\label{fig:PPara}Temperature dependence of the frequency and HWHM for the four Raman-active phonon modes of the In2 site of CeCoIn$_5$. The solid lines represent the fits to the anharmonic decay model [Eqs.\,(\ref{eq:energyTwo}) and (\ref{eq:gammaTwo})].}
\end{figure}

In Fig.~\ref{fig:PPara} we show the temperature dependence of both frequency and half width at half maximum (HWHM) for the Raman-active phonon modes. These spectral parameters are obtained by fitting the phonon modes with a Lorentzian lineshape.

For the two $E_{g}$ modes, the temperature dependence of both frequency $\omega_p(T)$ and HWHM $\gamma_p(T)$ can be accounted for by a standard model assuming anharmonic decay into two phonons with identical frequencies and opposite momenta~\cite{Klemens1966}:
\begin{equation}
\omega_p(T)=\omega_0-\omega_2[1+\frac{2}{e^{\hbar\omega_0/2k_B T}-1}],
\label{eq:energyTwo}
\end{equation}
and
\begin{equation}
\gamma_p(T)=\gamma_0+\gamma_2[1+\frac{2}{e^{\hbar\omega_0/2k_B T}-1}].
\label{eq:gammaTwo}
\end{equation}
In this model, the electronic contribution to phononic self energy is incorporated into the constants $\omega_0$ and $\gamma_0$. The parameters obtained from the anharmonic decay fits are summarized in Table~\ref{tab:Decay}.

\begin{table}
\caption{\label{tab:Decay}The parameters obtained from the anharmonic decay fits [Eqs.\,(\ref{eq:energyTwo}) and (\ref{eq:gammaTwo})]. All units are in cm$^{-1}$.}
\begin{ruledtabular}
\begin{tabular}{ccccc}
Mode&$\omega_0$&$\omega_2$&$\gamma_0$&$\gamma_2$\\
\hline
A$_{1g}$&164.5(1)&0.57(1)&1.54(6)&0.13(2)\\
B$_{1g}$&153.4(1)&0.39(3)&1.94(19)&0.35(6)\\
E$_{g}^{(1)}$&65.98(4)&0.173(6)&0.76(3)&0.027(4)\\
E$_{g}^{(2)}$&128.1(1)&0.42(2)&1.62(6)&0.067(18)\\
\end{tabular}
\end{ruledtabular}
\end{table}

For the $A_{1g}$ and $B_{1g}$ modes below T*, both frequency and HWHM become smaller than the value predicted by the anharmonic decay model. Similar anomaly has been observed for the $A_{1g}$ mode by Raman~\cite{Raman2007} and ultrafast optical spectroscopy~\cite{Fast2020}. However, we note that the resonance mode dominated by the scalar moduli, measured by resonant ultrasound spectroscopy, exhibits no anomaly across T*~\cite{Ramshaw2015}.

The anomalous behavior for the $A_{1g}$ and $B_{1g}$ modes indicates that the constants $\omega_0$ and $\gamma_0$ in Eqs.\,(\ref{eq:energyTwo}) and (\ref{eq:gammaTwo}), and in turn the electronic contribution to phonon self energy, become noticeably temperature dependent for these two modes below T*. We relate such anomaly to the development of Kondo-lattice coherence: formation of the heavy-quasiparticle band enhances the density of states near the Fermi level. Such change from incoherent to coherent electronic states below T* influences the electronic contribution to the phononic self energy. The experimental results indicate that the increase of coherent spectral weight at Fermi level results in softening phonon frequency and reduced electron-phonon scattering rate.

The lattice vibrations cause dynamic modulations of the electrostatic potential at Ce sites, and in turn the charge distribution of the CF states. As the $A_{1g}$ and $B_{1g}$ modes involve motion toward and away from the Ce sites, these two modes are significantly influenced by the coherence process and exhibit anomalous behaviors. In contrast, the two $E_{g}$ modes involve in-plane motion and show essentially no anomaly.

\section{Conclusions\label{sec:Con}}

In summary, we show by inelastic light scattering the effect of Kondo-lattice coherence on the electronic and phononic excitations for CeCoIn$_5$. Below the coherent temperature T*, we observe a suppression of low-frequency electronic continuum and anomalous temperature dependence of the phonon spectral parameters.

For the electronic excitations, the continuum in the $B_{2g}$ symmetry channel is suppressed below T*, with the Raman response following a cubic power law up to 50\,cm$^{-1}$. In contrast, the low-frequency continuum in the $B_{1g}$ channel shows no suppression across T*. As the electronic continuum results from the large electron self energy, the suppression of such continuum at low frequency indicates reduced electron-electron scattering rate. 
By considering the quasi-particle momentum dependence of the light-scattering vertex for these two symmetry channels, we find that the $\alpha$ and $\beta$ bands become coherent below T*, whereas the $\gamma$ band remains largely incoherent down to 10\,K.

For the phononic excitations, the temperature dependence of both frequency and linewidth of the $A_{1g}$ and $B_{1g}$ phonon modes deviates from anharmonic-decay behavior below T*. The development of coherent spectral weight near the Fermi level influences the phonon self energy. In particular, the coherence process softens phonon frequency and reduces electron-phonon scattering rate.

\begin{acknowledgments}
M.Y. and H.-H.K. contributed equally to this work. We thank Ruiqi Zhang for discussing the phononic excitations. 
The spectroscopic work at Rutgers (M.Y., H.-H.K., and G.B.) was supported by NSF Grants DMR-1709161 and DMR-2105001. 
The sample synthesis at Los Alamos (P.F.S.R. and E.D.B.) was supported by DOE Office of Basic Energy Sciences, Division of Materials Science and Engineering.
The work at NICPB was supported by the European Research Council (ERC) under Grant Agreement No. 885413.
\end{acknowledgments}


%

\end{document}